\documentclass[twocolumn,trackchanges]{aastex62}

\hypersetup{linkcolor=red,citecolor=blue,filecolor=cyan,urlcolor=magenta}
\usepackage{natbib}
\usepackage{amsthm}
\usepackage{amssymb}
\usepackage{amsmath}
\usepackage{bm}

\received{\today}
\revised{, 2020}
\accepted{, 2020}
\submitjournal{ApJ}

\shorttitle{Proton firehose instability in the solar wind}
\shortauthors{Shaaban et al.}

\begin{document}

\title{A new low-beta regime for unstable proton firehose modes in bi-Kappa distributed plasmas}

\correspondingauthor{S.M.Shaaban}
\email{s.m.shaaban88@gmail.com}

\author[0000-0003-0465-598X]{S.M.Shaaban}
\affil{Institute of Experimental and Applied Physics, University of Kiel, Leibnizstrasse 11, D-24118 Kiel, Germany}
\affiliation{Theoretical Physics Research Group, Physics Dept., Faculty of Science, Mansoura University, 35516, Mansoura, Egypt}

\author[0000-0002-8508-5466]{M. Lazar}
\affiliation{Centre for Mathematical Plasma Astrophysics, KU Leuven, Celestijnenlaan 200B, B-3001 Leuven, Belgium}
\affiliation{Institut f\"ur Theoretische Physik, Lehrstuhl IV: Plasma-Astroteilchenphysik, Ruhr-Universit\"at Bochum, D-44780 Bochum, Germany}

\author[0000-0002-7388-173X]{R. F. Wimmer-Schweingruber}
\affiliation{Institute of Experimental and Applied Physics, University of Kiel, Leibnizstrasse 11, D-24118 Kiel, Germany}

\author[0000-0002-9151-5127]{H. Fichtner}
\affiliation{Institut f\"ur Theoretische Physik, Lehrstuhl IV: Weltraum- und Astrophysik, Ruhr-Universit\"at Bochum, D-44780 Bochum, Germany}

\begin{abstract}
In the solar wind plasma an excess of kinetic temperature along the background magnetic field stimulates proton firehose modes to grow if the parallel plasma beta parameter is sufficiently high, i.e., $\beta_{p \parallel}\gtrsim 1$. This instability can prevent the expansion-driven anisotropy from increasing indefinitely, and explain the observations. Moreover, such kinetic instabilities are expected to be even more effective in the presence of suprathermal Kappa-distributed populations, which are ubiquitous in the solar wind, are less affected by collisions than the core population, but contribute with an additional free energy. In this work we use both linear and extended quasi-linear (QL) frameworks to characterize the unstable periodic proton firehose modes (propagating parallel to the magnetic field) under the influence of suprathermal protons. Linear theory predicts a systematic stimulation of the instability, suprathermals amplifying the growth rates and decreasing the instability thresholds to lower anisotropies and lower plasma betas ($\beta_{p \parallel}<1$). In perfect agreement with these results, the QL approach reveals a significant enhancement of the resulting electromagnetic fluctuations up to the saturation with a stronger back reaction on protons, leading also to a faster and more efficient relaxation of the temperature anisotropy.
\end{abstract}

\keywords{solar wind --- plasmas --- instabilities --- waves }
\section{Introduction}\label{Sec.1}
%
In heliospheric plasmas particle-particle collisions are rare and therefore particle velocity distribution functions (VDFs) can exhibit non-thermal features which are measured in-situ, such as the suprathermal populations and deviations from isotropy \citep{Marsch2006, Stverak2008}. The most plausible scenarios for the formation and preservation of such populations suggest an important role of the small-scale plasma waves and fluctuations. Thus, suprathermal particles can be sustained by a certain level of wave turbulence, but, at the same time, the same suprathermals can stimulate both the spontaneous and induced emissions \citep{Lazar2018Spont, Kim2018, Lazar2019, ShaabanApJ2019, Shaaban2021}. On the other hand, kinetic instabilities driven by the anisotropic populations lead to enhanced fluctuations which should act back on particles triggering their relaxation below the instability thresholds \citep{ShaabanMNRAS2021}. 
Even for the low-energy (core) populations, still influenced by collisions (given their Maxwellian velocity distributions (VDs) and collisional age), the limits of temperature anisotropies reported by solar wind observations appear to be bounded by the instability thresholds \citep{Kasper2002, Stverak2008}. The adiabatic expansion of the solar wind \citep{Chew1956, Matteini2007} predicts at 1~AU a significant excess of parallel temperature of protons $T_\parallel \gg T_\perp$ ($\parallel$ and $\perp$ denoting directions parallel and perpendicular to the magnetic field), but the observed quasi-stable states are below the anisotropy thresholds of proton firehose instabilities \citep{Matteini2007, Kasper2003, Bale2009, Michno2014, Huang2020}.

When described by an idealized bi-Maxwellian distribution function the anisotropic protons (subscript $p$) with $T_{p \parallel}>T_{p \perp}$ may destabilize two distinct branches of proton firehose instabilities \citep{Gary1993, Maneva2016, Hunana2017}. The periodic proton firehose
(PFH) instability, with a non-zero real frequency $\omega_r\neq 0$, evolves from a right handed (RH) polarized mode with dominant growth rate for parallel propagation, i.e., $\bm{k} \times \bm{B}_0=\bm{0}$, while the aperiodic ($\omega_r = 0$) proton firehose instability evolves only for oblique angles, i.e., $\bm{k} \times \bm{B}_0\neq \bm{0}$ \citep{Gary1993, Maneva2016, Hunana2017}. 
Close to marginal stability (low growth rates) the PFH instability evolves in general faster with growth rates higher than the aperiodic mode \citep{Gary1993, Hellinger2006, Maneva2016}, suggesting a primary role in enhancing the electromagnetic fluctuations which can scatter the protons and limit their anisotropy. 
In this case the destabilization of both proton firehose branches is conditioned by a sufficiently high parallel plasma beta $\beta_{p \parallel}\gtrsim 1$, see for instance \citet{Maneva2016} and references therein. The parallel plasma beta $\beta_{p \parallel}= p_{\rm kin}/p_{\rm mag}$ is the ratio of the parallel kinetic pressure $p_{\rm kin} =n_p k_B T_{p \parallel}$ to the magnetic pressure $p_{\rm mag}=B_0^2/8\pi$ (with the proton number density $n_p$, the proton parallel temperature $T_{p \parallel}$, the magnetic field $B_0$, and Boltzmann constant $k_B$). The plasma beta parameter can discriminate between plasma regimes dominated either by the magnetic effects when $\beta < 1$, or by the kinetic effects of plasma particles when $\beta > 1$. In the present paper we will show that such a condition for the excitation of proton firehose instabilities, i.e., $\beta_{p \parallel} \gtrsim 1$, can change considerably in the presence of suprathermal protons.

Indeed, the way suprathermals are involved in the excitation of proton firehose instabilities is not yet clear, as these populations are not affected by collisions, and require a different approach. In space plasmas the VDs of protons and heavier ions exhibit high-energy tails, formed by the suprathermal populations and well described by the (bi-)Kappa distribution functions \citep{Christon1989, Collier1996, Tylka2006, Pierrard2010, Ebert2012, Christon2017, Yu2017, Lario2019}. Ions with suprathermal distributions have been observed in the solar wind \citep{Christon1989, Collier1996}, coronal mass ejections (CMEs) \citep{Tylka2006}, co-rotating interaction regions (CIRs) \citep{Ebert2012, Yu2017}, magnetosphere \citep{Christon2017}, and interplanetary shocks \citep{Lario2019}. Suprathermal populations are expected to contribute with an additional free energy that may stimulate kinetic instabilities and enhance the wave fluctuations. A series of recent studies have confirmed this stimulating effect on spontaneous emissions \citep{Lazar2018Spont}, and induced emissions as well, e.g., instabilities of whistler waves \citep{Lazar2019} and electromagnetic ion-cyclotron modes \citep{Shaaban2016Supra, Shaaban2021}.  

However, similar studies of PFH instabilities have led to less systematic results. Thus, in the case of the periodic branch, also known as the parallel PFH instability, it is suggested that suprathermal protons may either stimulate the instability if driven by small anisotropies, or inhibit it when excited by large anisotropies \citep{Lazar2011, Astfalk2016}. The other, aperiodic (or oblique) branch has a different behavior, at least for the low anisotropies described by \cite{Astfalk2016}, suprathermal ion populations lead to lower growth rates and, implicitly, a stabilization of the plasma.
These desultory results are due to a misinterpretation of suprathermal populations and their implications by adopting a simplified Kappa distribution, which enables only a contrast with a Maxwellian limit of same temperature, and, thus, it cannot reveal and quantify the effects of suprathermals \citep{Lazar2015Destabilizing,LazarAA2016}. 
An analysis able to highlight these populations and their effects on our PFH instability should be based on the contrast between bi-Kappa distributed protons (with tails enhanced by the suprathermals) and the bi-Maxwellian quasi-thermal core of the distribution, without the suprathermal tails\footnote{For more detailed explanations see \cite{Lazar2015Destabilizing, LazarAA2016}}.

In the present paper we adopt such a realistic approach to describe both linearly and quasi-linearly (QL) the periodic PFH modes driven by bi-Kappa protons. For low and moderate plasma beta regimes, like the ones of interest in our study, the periodic PFH instability may evolve faster and dominate the aperiodic branch \citep{Hellinger2006,Astfalk2016}.  
QL analysis enables us to investigate not only the saturation of growing fluctuations but, also, their back reaction on protons, contributing to their relaxation. 
The velocity moment-based QL theory applied here offers a reliable and straightforward description of the main features of kinetic instabilities driven by the temperature anisotropy of plasma particles \citep{Davidson1968, Yoon2017}. Numerical simulations confirm the major role of the main velocity moments, such as temperature components parallel and perpendicular to the magnetic field, and also show that transient deformations of the distributions fade over time, and the initial shape (e.g., bi-Maxwellian, or bi-Kappa) is mainly restored by the extended time of relaxation \citep{Seough2014, Seough2015, Yoon2017PoP, Lazar2018JGR, Micera2020}\footnote{Other QL diffusion theories intended to reproduce transient deformations of the anisotropic distribution (e.g., \cite{Jeong2020} may be very complicated and still limited to a number of approximations. Their implementation to fully describe the evolution of instability, including saturation of EM fluctuations and relaxation of the distribution, is not yet feasible.}.
In Section~\ref{Sec.2} we first introduce the anisotropic bi-Kappa model for protons, while the electrons are assumed Maxwellian and initially isotropic. Then we provide linear and QL equations used to describe the unstable PFH solutions. Numerical solutions are derived and discussed in Section \ref{Sec.3}, analyzing in detail the effects of suprathermal protons. The results of the present work are summarized in Section \ref{Sec.4}.

\section{Theoretical approaches}\label{Sec.2}
%

\subsection{Linear formalism}
We assume a collisionless and homogeneous plasma, in the initial configuration with bi-Kappa distributed protons (subscript $p$) and Maxwellian isotropic electrons (subscript $e$). The bi-Kappa distribution function \citep{Lazar2015Destabilizing}
\begin{align}\label{eq1}
f_{\kappa,p}\left(v_{\parallel }, v_{\perp }\right)=&\frac{1}{\pi ^{3/2} \theta_{p \parallel} \theta_{p \perp}^{2}}
\frac{\Gamma\left( \kappa +1\right)}{\kappa^{3/2}\Gamma \left( \kappa -1/2\right)}\nonumber\\
&\times \left[ 1+\frac{v_{\parallel }^{2}}{\kappa~\theta_{p \parallel }^{2}}
+\frac{v_{\perp }^{2}}{\kappa~\theta_{p \perp}^{2}}\right] ^{-\kappa-1},
\end{align}
is defined in terms of the normalization velocities $\theta_{p \parallel,  \perp}(t)$, varying with time ($t$) in our QL analysis, and related to the anisotropic temperature components, as given by the second order moments 
\begin{align}\label{eq2}
T_{p \parallel}^{\kappa}=\frac{2\kappa}{2\kappa-3}\frac{m_p \theta^2_{p \parallel}}{2 k_B}, \text{ and } T_{p \perp}^{\kappa}=\frac{2\kappa}{2\kappa-3}\frac{m_p \theta^2_{\perp p}}{2 k_B}
\end{align}
\textcolor{black}{In the absence of suprathermal particles, i.e., in the limit of $\kappa \rightarrow~\infty$, the distribution reduces to that of the core population described (approximately) by the following bi-Maxwellian  \citep{Lazar2015Destabilizing, LazarAA2016}
\begin{align}\label{eq3}
f_{M,p}\left( v_{\parallel },v_{\perp }\right) =&\frac{1}{\pi
^{3/2} \theta_{p \parallel} \theta_{p \perp}^{2}}\exp \left(
-\frac{v_{\parallel }^{2}} {\theta_{p \parallel}^{2}}-\frac{v_{\perp
}^{2}}{\theta_{p \perp}^{2}}\right),   
\end{align}
with temperature components 
\begin{align}\label{Max-T}
T_{p \parallel}=\frac{m_p \theta^2_{p \parallel}}{2 k_B}< T_{p \parallel}^{\kappa}, \text{ and } T_{p \perp}=\frac{m_p \theta^2_{\perp p}}{2 k_B} < T_{p \perp}^{\kappa},
\end{align}
which are obtained from Eq.\eqref{eq2} in the same limit of $\kappa \to \infty$.
The core temperatures are lower (the core is in general much cooler) by a factor $\alpha^2= \kappa/ (\kappa-3/2) > 1$, and $\theta_{p \perp, \parallel}(t)=\sqrt{2k_B T_{\parallel,\perp p}(t)/m_p}$ become the associated (well-defined) thermal velocities. 
}

We assume a quasi-neutral proton-electron plasma $n_p \approx n_e$ and, to isolate and describe only the effects of the suprathermal protons on PFH instability, initially (at $t=0$) the electrons (subscript $e$) are considered isotropic Maxwellian, similar to Eq.~\eqref{eq3} but with $\theta_{e,\parallel} =\theta_{e,\perp} = \theta_e$). For such a plasma system, proton firehose modes propagating parallel to the background magnetic field (i.e., ${\bm k} \times {\bm B}_0=\bm{0}$) are described by the following dispersion relation \citep{Shaaban2017}
\textcolor{black}{%
\begin{align} \label{dis}
\frac{c^2k^2}{\omega^2_{pp}}=&\frac{\omega^2_{pe}}{\omega^2_{pp}} \left[A_e-1+\frac{A_e~\omega -\left(A_e-1\right)\Omega_e}{k \theta_{e \parallel}} Z_{M, e}\left(\xi_e^{-}\right)\right]\nonumber\\
&+A_p-1+\frac{A_p~\omega +\left(A_p-1\right)\Omega_p}{k \theta_{p \parallel}}Z_{\kappa,p}\left(\xi_p^{+}\right),
\end{align} 
where $k$ is the wave-number, $c$ is the light speed, $\omega_{p j}=~\sqrt{4\pi n_j e^2/m_j}$ and $\Omega_j=~e B_0/m_j c$ are the plasma frequency and the non-relativistic gyro-frequency of the plasma species $j$, $\omega=\omega_r+~i\gamma$ is the wave frequency, $A_j\equiv \theta^2_{j \perp}/\theta^2_{j \parallel} \equiv T_{j \perp}^\kappa/T_{j \parallel}^\kappa \equiv ~T_{j \perp}/T_{j \parallel}$ is the temperature anisotropy of the plasma species $j$,
\begin{align} \label{eq5}
 Z_{\kappa,p}\left( \xi_{p}^{+}\right) =&\frac{1}{\pi ^{1/2}\kappa^{1/2}}\frac{\Gamma \left( \kappa \right) }{\Gamma \left(\kappa -1/2\right) }\nonumber\\
     &\int_{-\infty }^{\infty }\frac{\left(1+x^{2}/\kappa \right) ^{-\kappa}}{x-\xi_{p}^{+}}dx,\  \Im \left(\xi _{p}^{+}\right) >0.
\end{align}
is the modified dispersion function for Kappa-distributed plasmas \citep{Lazar2008} of argument $\xi_p^{+}=~(\omega+\Omega_p)/(k \theta_{p \parallel})$,
and
\begin{equation}  \label{eq6}
Z_{M,e}\left( \xi_{e}^{-}\right) =\frac{1}{\sqrt{\pi}}\int_{-\infty
}^{\infty }\frac{\exp \left(-x^{2}\right) }{x-\xi _{e}^{-}}dx,\
\ \Im \left( \xi_{e}^{-}\right) >0, 
\end{equation}
is the plasma dispersion function \citep{Fried1961} of argument $\xi_e^{-}=(\omega-|\Omega_e|)/(k \theta_{e \parallel})$.}

\textcolor{black}{The dispersion relation \eqref{dis} can be rewritten with normalized quantities as follows
\begin{align} \label{eq4}
\tilde{k}^2&=\mu \left[A_e-1+\frac{A_e~\tilde{\omega} -\left(A_e-1\right)\mu}{\tilde{k} \sqrt{\mu \beta_{e \parallel}}} Z_{M, e}\left(\frac{\tilde{\omega}-\mu}{\tilde{k} \sqrt{\mu\beta_{e \parallel}}}\right)\right]\nonumber\\
&+A_p-1+\frac{A_p~\tilde{\omega} +\left(A_p-1\right)}{\tilde{k} \sqrt{\beta_{p \parallel}}}Z_{\kappa,p}\left(\frac{\tilde{\omega}+1}{\tilde{k} \sqrt{\beta_{p \parallel}}}\right),
\end{align} 
where $\tilde{k}=ck/\omega_{p p}$ is the normalized wave-number, $\tilde{\omega}=~\omega/\Omega_p$ is the normalized wave frequency, $\beta_{p \parallel}=\theta_{p \parallel}^2 ~\omega_{pp}^2/(c^2~\Omega_p^2)$ and $\beta_{
e \parallel}=\theta_{e \parallel}^2~ \omega_{pe}^2/(c^2~\Omega_e^2)$ are, respectively, the parallel plasma beta parameters of protons and electrons, $\mu=m_p/m_e$ is the proton to electron mass ratio.}

\textcolor{black}{In order to outline the effects of suprathermal protons, we compare unstable solutions of Eq.~(\ref{eq4}) with the ones obtained for the Maxwellian core (in the absence of suprathermals) 
\begin{align} \label{eq4-M}
\tilde{k}^2&=\mu \left[A_e-1+\frac{A_e~\tilde{\omega} -\left(A_e-1\right)\mu}{\tilde{k} \sqrt{\mu \beta_{e \parallel}}} Z_{M, e} \left(\frac{\tilde{\omega}-\mu}{\tilde{k} \sqrt{\mu\beta_{e \parallel}}}\right)\right]\nonumber\\
&+A_p-1+\frac{A_p~\tilde{\omega} +\left(A_p-1\right)}{\tilde{k} \sqrt{ \beta_{p \parallel}}} Z_{M,p}\left(\frac{\tilde{\omega}+1}{\tilde{k} \sqrt{\beta_{p \parallel}}}\right),
\end{align} 
where $Z_{M,p}$ is a Maxwellian dispersion function, similar to Eq.~(\ref{eq6}), but with a proton argument $\xi_p^+$. This equation is straightforwardly obtained from Eq.~(\ref{eq4}), in the limit $\kappa \to \infty$.}     

\textcolor{black}{At this point, we should emphasize a few aspects to avoid confusion, and motivate our analysis in the next sections. 
The expressions of the proton argument in Eqs.~\eqref{eq4} and \eqref{eq4-M} are the same. However, the frequencies and wave numbers of the unstable solutions 
are not the same, being implicitly modified in the presence of suprathermals by the Kappa dispersion function $Z_{\kappa,p}$, through the (finite) power exponent $\kappa$ (see also the footnote \footnote{The same dispersion relation \eqref{eq4} can be expressed in terms other, slightly modified Kappa dispersion functions, e.g., \cite{Summers1991}, leading to expressions depending explicitly on the $\kappa$ parameter, but this will not affect the wave (unstable or stable) solutions \citep{Lazar2008}.}). This already suggests that to highlight the effects of suprathermal protons we can compare the solutions of Eqs.~\eqref{eq4} and \eqref{eq4-M} obtained for the same beta value, specific to the (bi-)Maxwellian core of our (bi-)Kappa distribution.}

\subsection{Quasi-linear (QL) formalism}
In a QL formalism, the temporal evolution of the VDFs $f_j$ of the plasma species $j$ is described by the general kinetic equation in the diffusion approximation \citep{Yoon2017}
\begin{align} \label{eq7}
\frac{\partial f_j}{\partial t}&=\frac{i e^2}{4m_j^2 c^2~ v_\perp}\int_{-\infty}^{\infty} 
\frac{dk}{k}\left[ \left(\omega^\ast-k v_\parallel\right)\frac{\partial}{\partial v_\perp}+ 
k v_\perp\frac{\partial}{\partial v_\parallel}\right]\nonumber\\
&\times\frac{ v_\perp \delta B^2(k, \omega)}{\omega-kv_\parallel\pm \Omega_j}\left[ 
\left(\omega-k v_\parallel\right)\frac{\partial f_j}{\partial v_\perp}+ k v_\perp
\frac{\partial f_j}{\partial v_\parallel}\right], 
\end{align}
{where $\pm$ denote, respectively, the circular right-handed (RH) or left-handed (LH) polarization, and $ B^2(k)$ is the spectral magnetic wave-energy density of the enhanced fluctuations, which is described by the wave kinetic equation}
\begin{equation} \label{eq8}
\frac{\partial~\delta B^2(k)}{\partial t}=2~\Im(\omega) \delta B^2(k),
\end{equation}
{with the instantaneous growth rate $\Im(\omega)=\gamma$ of the PFH instability derived from the linear dispersion relation \eqref{eq4}.} 
%
\begin{figure*}[t!]
    \centering
    \includegraphics[trim= 3.cm 15.5cm 2.2cm 2.5cm, clip, width=\textwidth]{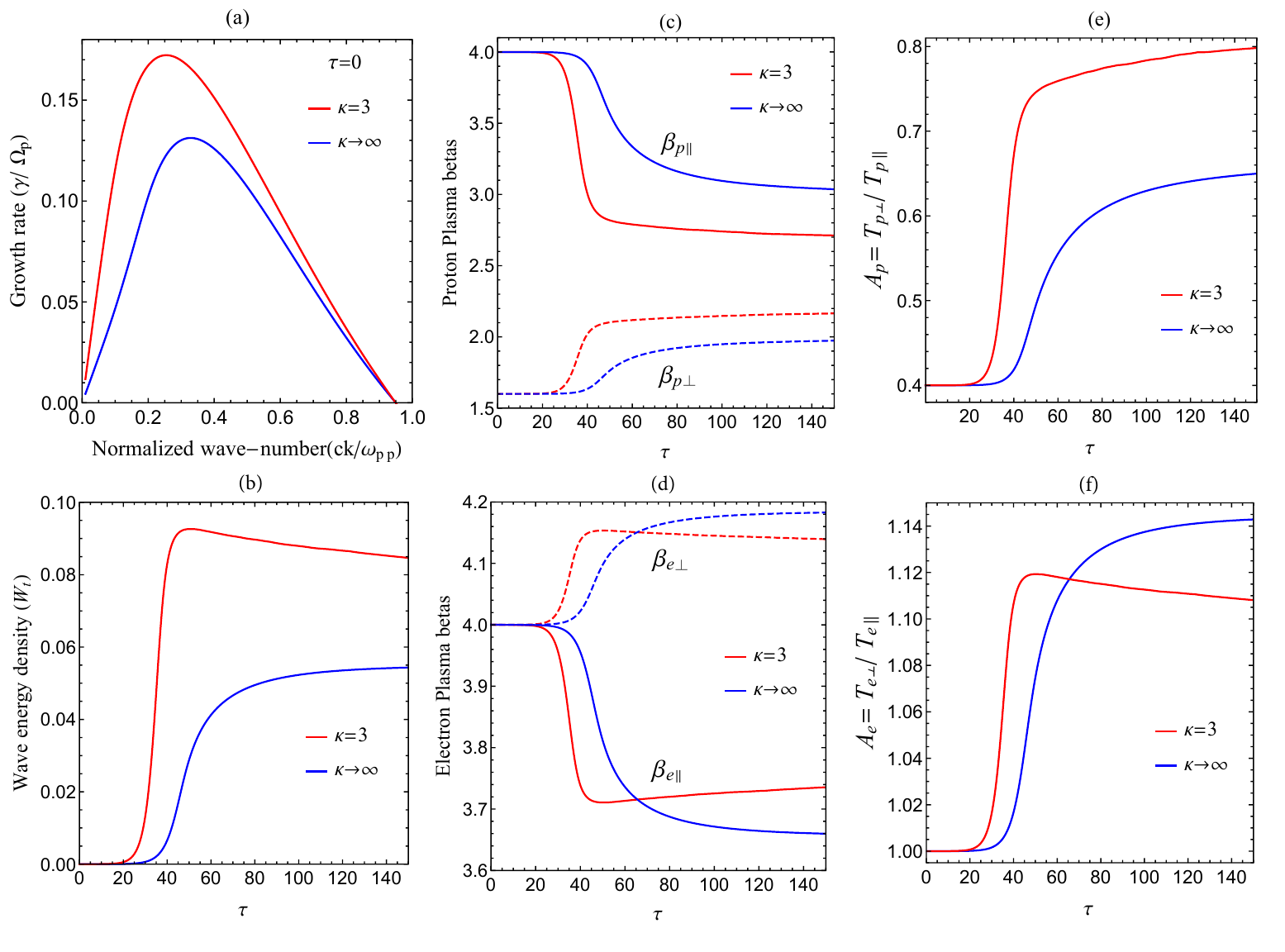}
    \caption{Numerical solutions of the PFH modes for $\kappa=3$ (red) and $\kappa\rightarrow \infty$ (blue): growth rates from linear theory (panel a), and QL temporal profiles of the wave energy density $W_t$ (panel b) and their back reactions on the beta parameters of protons 
    $\beta_{p \perp,\parallel}$ (panel c) and electrons $\beta_{e \perp,\parallel}$ (panel d) as well as their temperature anisotropies $A_p$ (panel e) and $A_e$ (panel f).}
    \label{fig:1}
\end{figure*}

\textcolor{black}{The QL kinetic equations for the time evolution of the temperature components defined as velocity moments (for protons and electrons) result from \eqref{eq7}, as follows
\begin{subequations}\label{eq9}
\begin{align}
\frac{dT^\kappa_{p \parallel}}{dt}&=\alpha^2\frac{dT_{p \parallel}}{dt}=\frac{m_p}{k_B}\frac{\partial}{\partial t}\int d^3v~ v_\parallel^2~f_{\kappa,p},\\
\frac{dT^\kappa_{p \perp}}{dt}&=\alpha^2\frac{dT_{p \perp}}{dt}=\frac{m_p}{2k_B}\frac{\partial}{\partial t}\int d^3v~ v_\perp^2~f_{\kappa,p},\\
\frac{dT_{e \parallel}}{dt}&=\frac{m_e}{k_B}\frac{\partial}{\partial t}\int d^3v~ v_\parallel^2~f_{M, e},\\
\frac{dT_{e \perp}}{dt}&=\frac{m_e}{2k_B}\frac{\partial}{\partial t}\int d^3v~ v_\perp^2~f_{M, e}.
\end{align}
\end{subequations}
For the sake of simplicity we can rewrite these dynamical equations \eqref{eq9} in terms of the dimensionless quantities
\begin{subequations}\label{eq10}
\begin{align}
\frac{d\beta_{p \parallel}}{d\tau}=&2\int\frac{d\tilde{k}}{\tilde{k}^2} W(\tilde{k}) \left[A_p~\tilde{\gamma}+G_{p \parallel}~\eta_p^{+}\right]/\alpha^2,\label{eq10a}\\
\frac{d\beta_{p \perp}}{d\tau}=&-\int\frac{d\tilde{k}}{\tilde{k}^2} W(\tilde{k})\left[\Lambda_p \tilde{\gamma}+ G_{p \perp}~\eta_p^{+}\right]/\alpha^2,\label{eq10b}\\
\frac{d\beta_{e \parallel}}{d\tau}=&2\int\frac{d\tilde{k}}{\tilde{k}^2} W(\tilde{k})\left[\mu A_e\tilde{\gamma}+ G_{e \parallel}~\eta_e^{-}\right],\\
\frac{d\beta_{e \perp}}{d\tau}=&-\int\frac{d\tilde{k}}{\tilde{k}^2} W(\tilde{k})\left[\mu\Lambda_e \tilde{\gamma}+ G_{e \perp}~\eta_e^{-}\right],
\end{align}
\end{subequations}
defining compactly
\begin{align*}
\tilde{\gamma}=&\gamma/\Omega_p,~~ W(\tilde{k})=\delta B^2(\tilde{k})/B_0^2,\\
\Lambda_j=&\left(2 A_j-1\right),~~ \tau=~\Omega_p~t,\\
\eta_p^{+}=&\left[A_p~\tilde{\omega}+\left(A_e-1\right)\right]Z_{p,\kappa}\left(\xi_p^{+} \right),\\
\eta_e^{-}=&\sqrt{\mu}\left[ A_e~\tilde{\omega}-\left(A_e-1\right)\mu\right]Z_{e}\left(\xi_e^-\right),\\
G_{p \parallel}=& \text{ Im} \frac{\tilde{\omega}+1}{\tilde{k}\sqrt{\alpha^2 \beta^\kappa_{p \parallel}}},~ G_{p \perp}= \text{Im} \frac{2i\tilde{\gamma}+1}{\tilde{k}\sqrt{\alpha^2 \beta^\kappa_{p \parallel}}},\\
G_{e \parallel}=& \text{ Im} \frac{\tilde{\omega}-\mu}{\tilde{k}\sqrt{\beta_{e \parallel}}},~ G_{e \perp}= \text{Im} \frac{2i\tilde{\gamma}-\mu}{\tilde{k}\sqrt{\beta_{e \parallel}}},
\end{align*}
}
and for the normalized spectral magnetic wave-energy density $W(\tilde{k})$
\begin{align}\label{e12}
\frac{\partial~W(\tilde{k})}{\partial \tau}=2~\tilde{\gamma}~ W(\tilde{k}).
\end{align}
\textcolor{black}{For a finite $\kappa$, these equations describe the time evolution of the instability induced by the bi-Kappa distributed protons. By contrast, in the absence of suprathermals, i.e., for $\kappa \to \infty$, the modified plasma dispersion function $Z_{p,\kappa}(\xi^+_p)$ converges to the standard dispersion function $Z_{p,M}(\xi^+_p)$ and $\alpha^2 \to 1$.  Consequently, we obtain a similar set of QL equations (not reproduced here), describing the instability triggered by the bi-Maxwellian core. 
}


\section{Numerical results}\label{Sec.3}
\begin{figure*}[t]
    \centering
    \includegraphics[trim= 3.cm 20.5cm 2.2cm 2.5cm, clip, width=\textwidth]{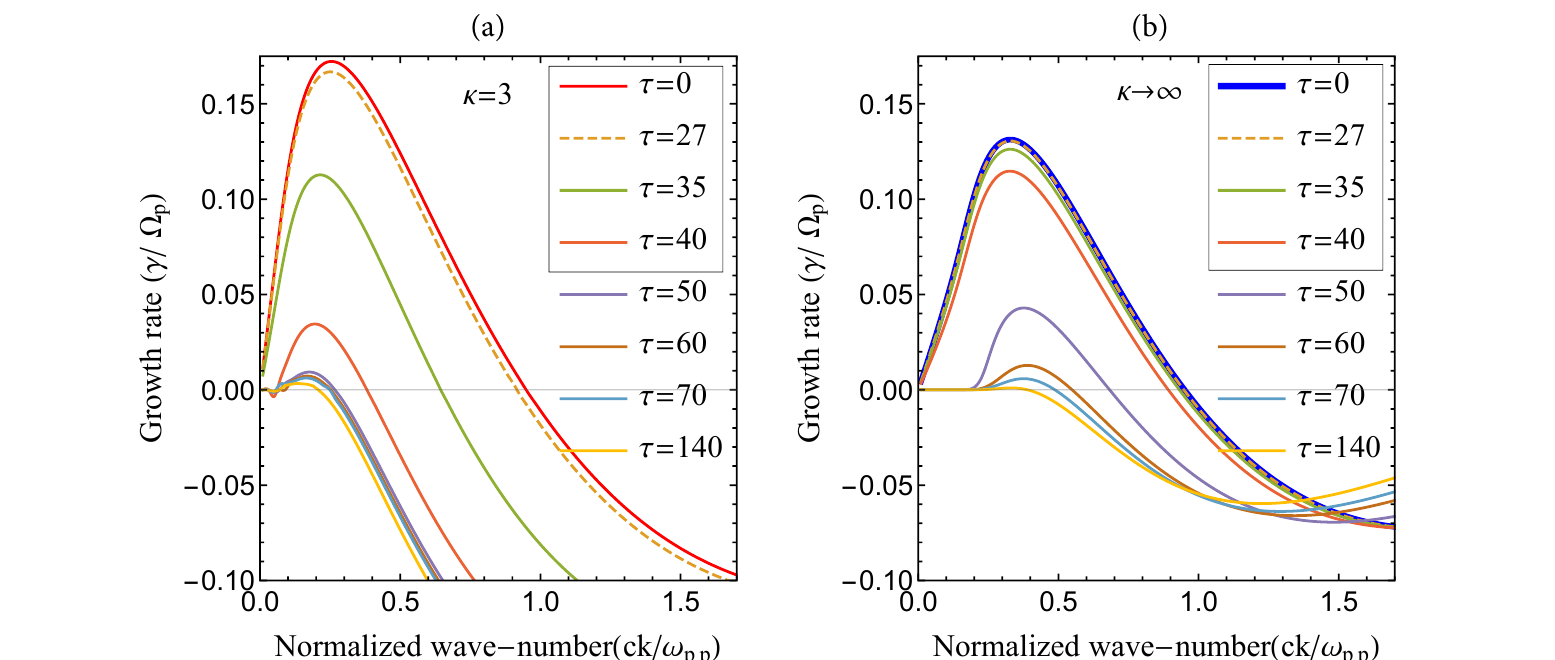}
    \caption{Instantaneous growth rates of the PFH instability for $\kappa=3$ (panel a) and $\kappa\rightarrow\infty$ (panel b).}
    \label{fig:2}
\end{figure*}
\begin{figure*}
    \centering
     \includegraphics[trim= 2.8cm 16.2cm 2.2cm 2.5cm, clip, width=\textwidth]{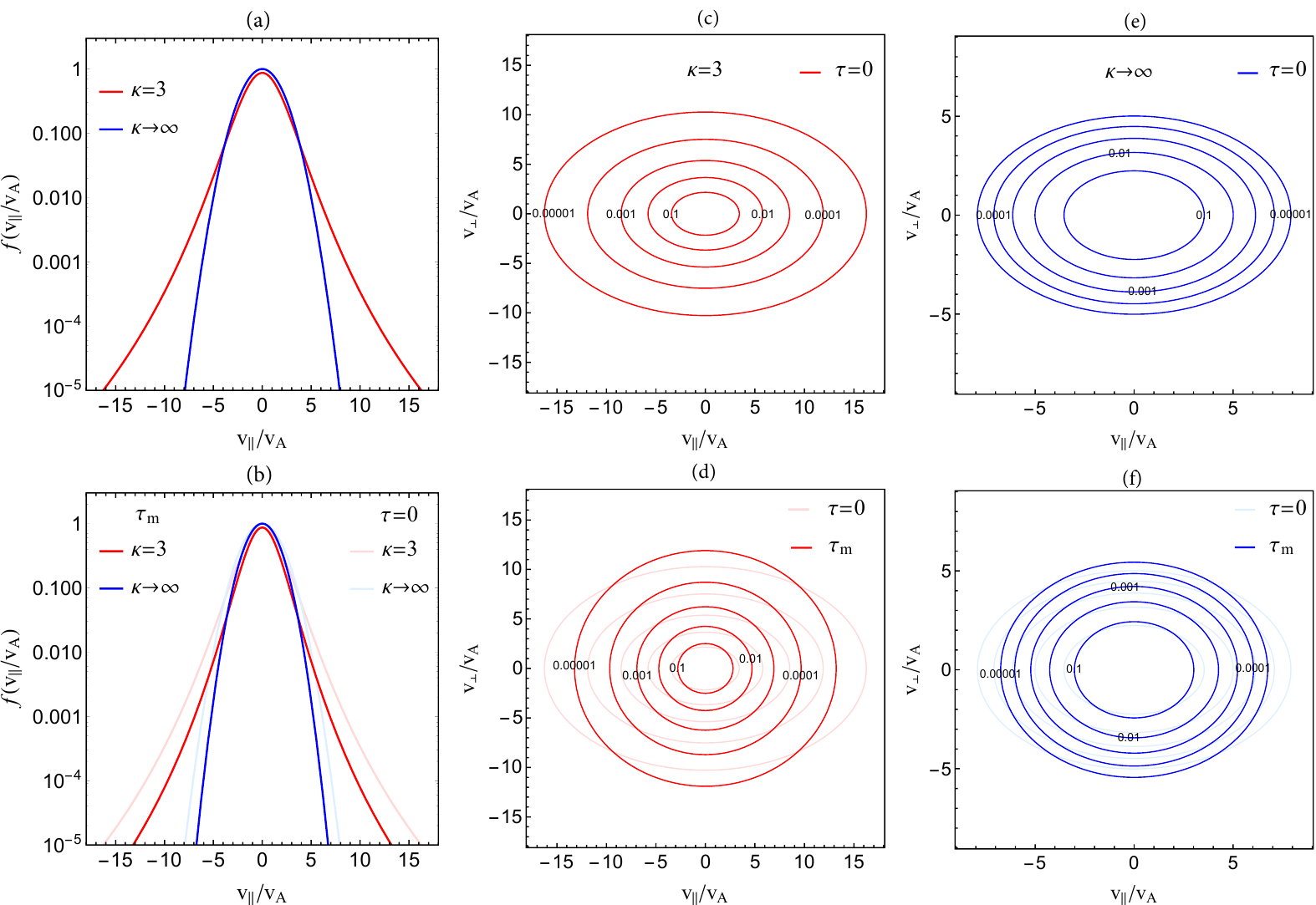}
    \caption{Parallel cuts (panels a and b) and contours (panels c, d, e, and f) of proton VDs, for $\kappa=3$ (red) and the bi-Maxwellian limit $\kappa \to\infty$ (blue), initially at $\tau=0$ (top) and after saturation $\tau=\tau_m$ (bottom).}
    \label{fig:3}
\end{figure*}
%
In this section we present the results of a comparative analysis, see Figures~\ref{fig:1}-\ref{fig:3}, contrasting the unstable PFH solutions obtained for bi-Kappa distributed protons, i.e., with $\kappa=3$, (red) with those obtained for the bi-Maxwellian limit $\kappa \to\infty$ (blue). Other initial plasma parameters, i.e., $A_p(0)=0.4$, $\beta_{p \parallel}(0)=4.0$,
$A_e(0)=1$, and $\beta_{e \parallel}(0)=4.0$, are fixed. 

Figure~\ref{fig:1} presents the numerical solutions of the PFH instability derived from the linear and QL approaches. Panel~(a) in Figure~\ref{fig:1} displays the growth rates of the PFH instability, markedly enhanced by the presence of suprathermal protons, i.e., for $\kappa=3$. The corresponding wave frequencies (not shown here) show minor variations with~$\kappa$. Beyond the linear dispersive properties of PFH instability, here we describe the QL increase of the instability, the temporal evolution of the enhanced magnetic wave-energy density of the PFH fluctuations $W_t=\int d\tilde{k}~\delta W(\tilde{k})$ (panel b), as well as their reactions back on the anisotropic protons and isotropic electrons. In panel~(b) the magnetic wave-energy density $W_t$ is markedly enhanced in the presence of the suprathermal protons ($\kappa=3$), confirming the linear theory  predictions in panel (a). For $\kappa=3$ the enhancement of the $W_t$ starts earlier, and $W_t$ shows a steeper growth profile before reaching a higher level of saturation.

The enhanced PFH fluctuations regulate the initial temperature anisotropy of protons $A_p\equiv \beta_{p \perp}/\beta_{p \parallel}<1$ (or  $A_p\equiv T_{p \perp}^\kappa/T_{p \parallel}^\kappa \equiv ~T_{p \perp}/T_{p \parallel}<1$) through cooling and heating processes reflected in panel (c) by, respectively, the parallel (solid lines) and perpendicular (dashed lines) plasma beta parameters. After saturation (i.e., at $\tau_{max}$) protons are less anisotropic (in  parallel direction) with $A_p(0)<A_p(\tau_{max})<1.0$, see panel (e). Initially isotropic, the electrons ($A_e(0)=1$) are subjected to parallel cooling (solid lines) and perpendicular heating (dashed lines), as shown in panel~(d) by the electron plasma beta parameters $\beta_{e \perp, \parallel}$. The electrons gain modest perpendicular anisotropy at later stages, i.e., $A_e(\tau_{max})\gtrsim 1.0$, see panel~(f). It is obvious that the enhanced fluctuations of PFH instability obtained in the presence of suprathermal protons ($\kappa=3$, red) lead to more pronounced effects on the proton plasma parameters than those obtained for a Maxwellian core only  ($\kappa\rightarrow\infty$), i.e., faster and stronger cooling and heating mechanisms for the plasma betas, and more efficient relaxation for the proton temperature anisotropy $A_p \to 1$, approaching the quasi-stable state after saturation, i.e., at $\tau_{max}$.

In Figure~\ref{fig:2} we plot the instantaneous growth rates of PFH instability at different time steps $\tau=~0, 27, 35, 40$, $50, 60, 70, 140$ (including the initial ones at $\tau=0$), for $\kappa=3$ (panel a) and $\kappa \to \infty$ (panel b). The instantaneous growth rates are plotted as a function of the normalized wave-number $ck/\omega_{p p}$. An interesting contrast is observed for the times $\tau=0, 27, 35, 40$, when suprathermals ($\kappa=3$) determine a quite severe change of the growth rates by comparison to those obtained for the bi-Maxwellian core ($\kappa \rightarrow \infty$). This explains  the enhanced PFH fluctuations in Figure~\ref{fig:1}, (panel b). For $\kappa \rightarrow \infty$ more significant changes are obtained only after $\tau=40$, when growth rates for $\kappa=3$ start to saturate.

In order to visualize the relaxation of the proton VDF under the effect of the enhanced PFH fluctuations we plot in Figure~\ref{fig:3} the normalized proton VDFs with suprathermals ($\kappa = 3$, red) and without them ($\kappa \rightarrow \infty$, blue). It is important to point out that in our QL analysis we assumed that $\kappa$ parameter does not change in time, assuming it constant. The shape of the initial VD changes only due to the variations of the main moments, such as temperature components. We show parallel cuts (panels a and b) and contours in velocity ($v_\perp/v_A, v_\parallel/v_A$)--space (panels c and d for $\kappa=3$, and  e and f for $\kappa \rightarrow \infty$), for the initial $\tau=0$ (top panels) and final time step $\tau=\tau_{max} $ after saturation (bottom panels). Here the VDF is normalized to the proton Alfv{\'e}n speed $v_A=B_0/\sqrt{4\pi n_p m_p}$, and for all panels the contour levels $10^{-1}, 10^{-2}, 10^{-3}, 10^{-4}$, and $10^{-5}$ of $f_{max}=1$ are shown. As one can see in the bottom panels, comparing to the initial state (indicated also by the light-red the light-red and light-blue contours in the background) the proton VDFs become less anisotropic in the parallel direction and more stable against PFH instability. As expected, the final state of the proton VDF for $\kappa=3$ is much less anisotropic than that obtained for $\kappa \rightarrow \infty$, and therefore more stable. It is worth noting that for $\kappa=3$ the kinetic (thermal) spread in the ($v_\perp/v_A, v_\parallel/v_A$)--space is two times wider than that for $\kappa \rightarrow \infty$.

Additional changes in the shape of the VD, as given by the variation of $\kappa$, are not easily captured in a QL approach. In a recent attempt to overcome this limitation, \cite{Moya2021} have proposed a new QL approach that includes the time variation of the $\kappa$ exponent during the relaxation of temperature anisotropy. However, this is only a zero-order approach needing further developments to be properly implemented in the QL theory.
%
\begin{figure}
    \centering
    \includegraphics[trim= 5.3cm 1.cm 4cm 2.5cm, clip, width=0.42\textwidth]{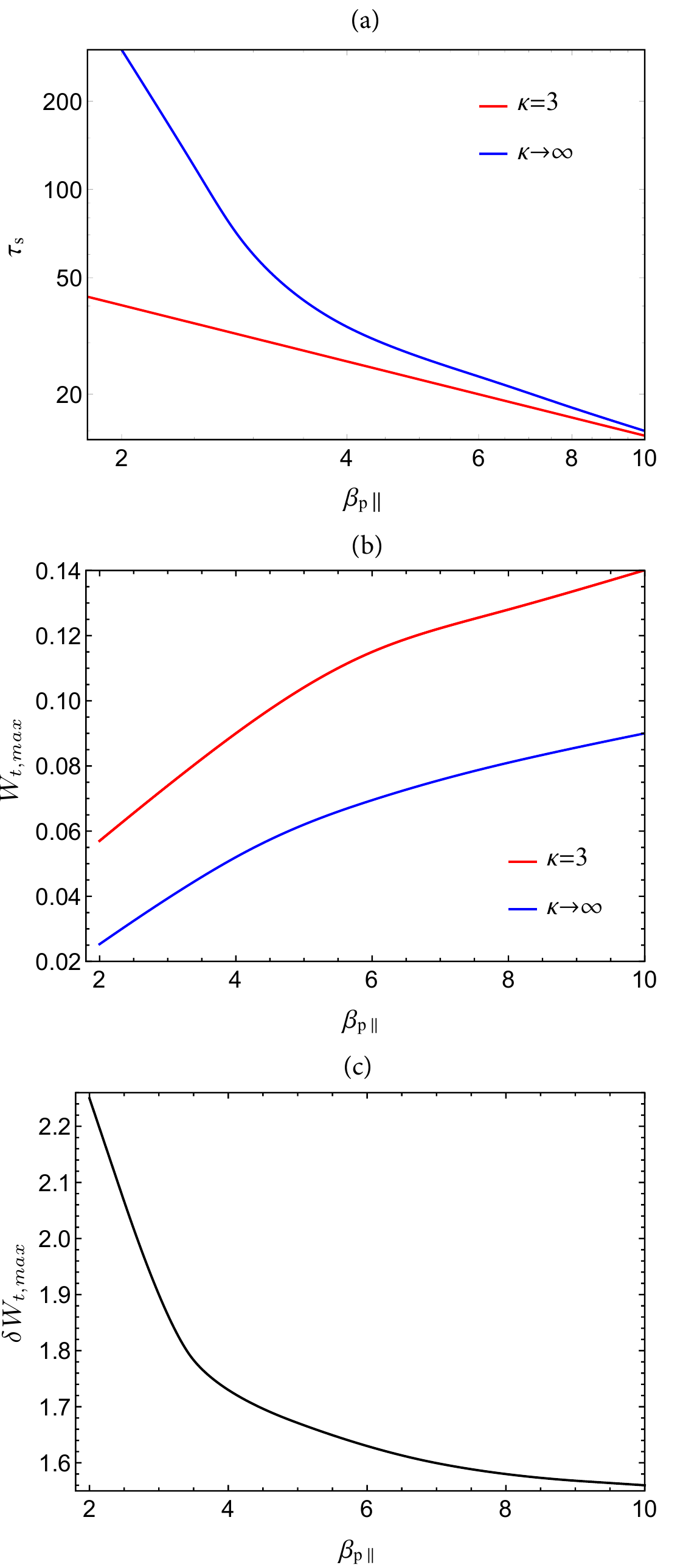}
    \caption{Effects of the proton power-index $\kappa=3$ (red) and $\kappa\rightarrow \infty$ (blue) on the starting time $\tau_s$ of the enhanced PFH fluctuations (panel a), and their maximum energy level  $W_{t, max}$ (panel b) as functions of $\beta_{p \parallel}$ for $2\leqslant\beta_{p \parallel}\leqslant10$. The ratio of these maximum levels $\delta W_{t, max}$ as a function of $\beta_{p \parallel}$ is shown in panel (c).}
    \label{fig:4}
\end{figure}
%

A more general perspective can be provided by studying the effects of the suprathermal protons on the QL development of the PFH instability as a function of the parallel plasma beta parameter of protons $\beta_{p \parallel}$. Figure~\ref{fig:4} presents comparisons for the magnetic wave-energy densities obtained for $\kappa=3$ (red) and $\kappa\rightarrow \infty$ (blue) for an extended range of plasma beta parameter $2.0~\leqslant~\beta_{p \parallel}\leqslant~10$, including the solar wind and planetary magnetosphere plasma conditions. \textcolor{black}{Panel (a) in Figure~\ref{fig:4} provides a comparison between the starting time $\tau_s$ of the enhancement of the PFH fluctuations as a function of $\beta_{p \parallel}$ for $\kappa=3$ (red) and $\kappa\rightarrow \infty$ (blue). Panel (a) shows that the enhancement of the PFH fluctuations start markedly earlier for $\kappa=3$ than those for $\kappa\rightarrow \infty$, especially for proton plasma beta parameter $\beta_{p \parallel}<4$. In general, the difference obtained for the starting time $t_s$ decreases with increasing $\beta_{p \parallel}$. Moreover, it is obvious that $t_s$ markedly decreases as $\beta_{p \parallel}$ increases, especially for the bi-Maxwellian limit $\kappa\rightarrow \infty$.}

\textcolor{black}{Panel (b) shows that the maximum magnetic wave-energy density $W_{t,max} (\tau)$ is markedly enhanced by increasing $\beta_{p \parallel}$. In the presence of the suprathermal protons (with $\kappa=3$)  $W_{t, max} (\tau)$ for $\beta_{p \parallel}=10$ is $\sim 2.5$ times higher than that for $\beta_{p \parallel}=2.0$, while in the Maxwellian limit ($\kappa\rightarrow \infty$) $W_{t, max} (\tau)$ for $\beta_{p \parallel}=10$ is $\sim 4.0$ times higher than that for $\beta_{p \parallel}=2.0$. The ratio between the maximum magnetic wave-energy density $W_{t, max}(\tau)$ for $\kappa=3$ and its Maxwellian limit $\kappa\rightarrow \infty$, which we name $\delta W_{t, max}(\tau)$ is displayed in panel (c) as a function of $\beta_{p \parallel}$. This ratio $\delta W_{t, max}(\tau)$ decreases with increasing $\beta_{p \parallel}$, starting from 2.25 and reaching a value of 1.55 at $\beta_{p \parallel}=10$. One conclusion to be drawn here is that the effects of the suprathermal populations on the enhanced PFH fluctuations significantly reduce with increasing the plasma beta parameter.}
%
\begin{figure}[t]
    \centering
   \includegraphics[trim= 5.3cm 0.3cm 4cm 2.5cm, clip, width=0.41\textwidth]{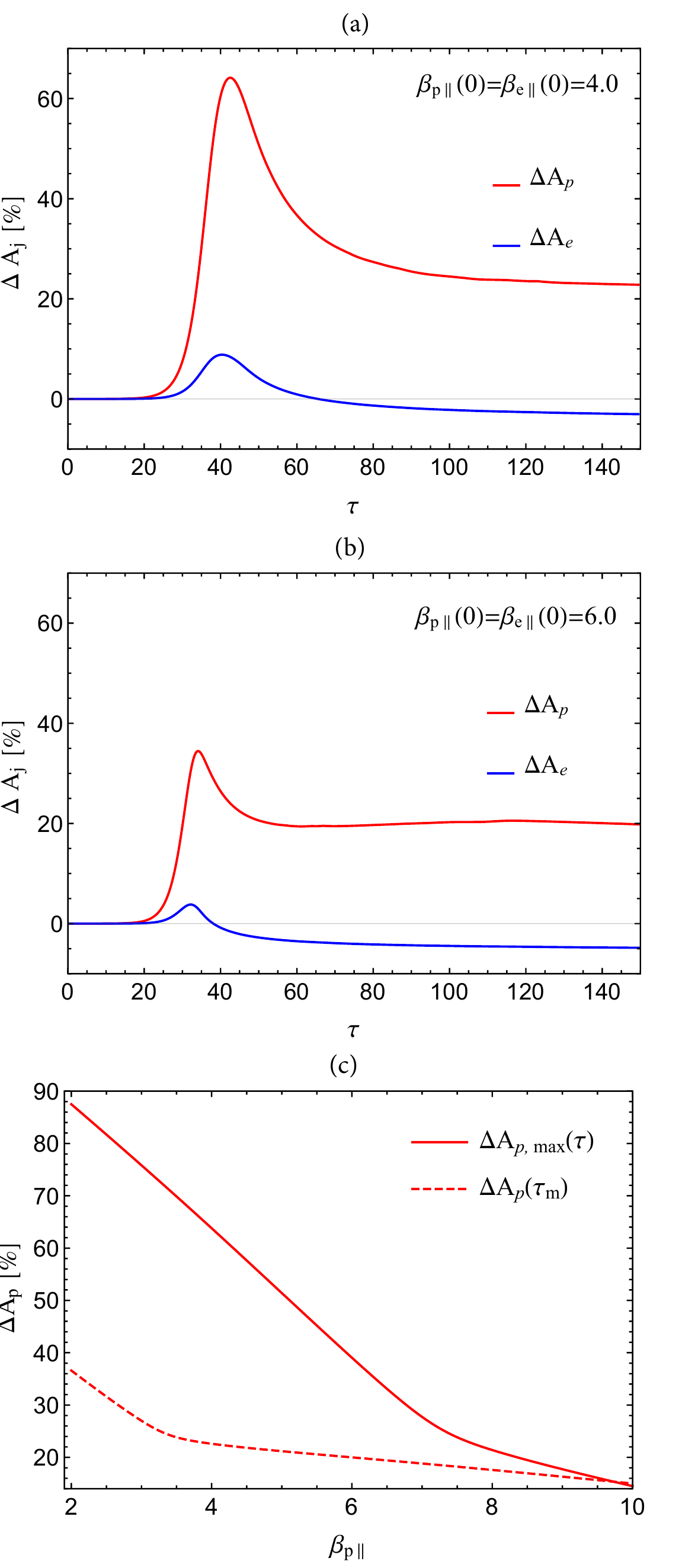}
    \caption{Temporal profiles of the relative percentage change of proton (red) and electron (blue) temperature anisotropies $\Delta A_j(\tau)= (A_j^{\kappa}(\tau)-A_j^{M}(\tau))/A_j^{M}(\tau) \times 100$ for $\beta_{p \parallel}=$4 (panel a), 6 (panel b). The maximum relative percentage difference $\Delta A_{p, max}(\tau)$ (red) and that obtained after saturation $\Delta A_{p}(\tau_m)$ (dashed red) are displayed in panel (c) as a function of $\beta_{p  \parallel}$.} 
    \label{fig:5}
\end{figure}
%
%
\begin{figure*}
    \centering
    \includegraphics[scale=0.48]{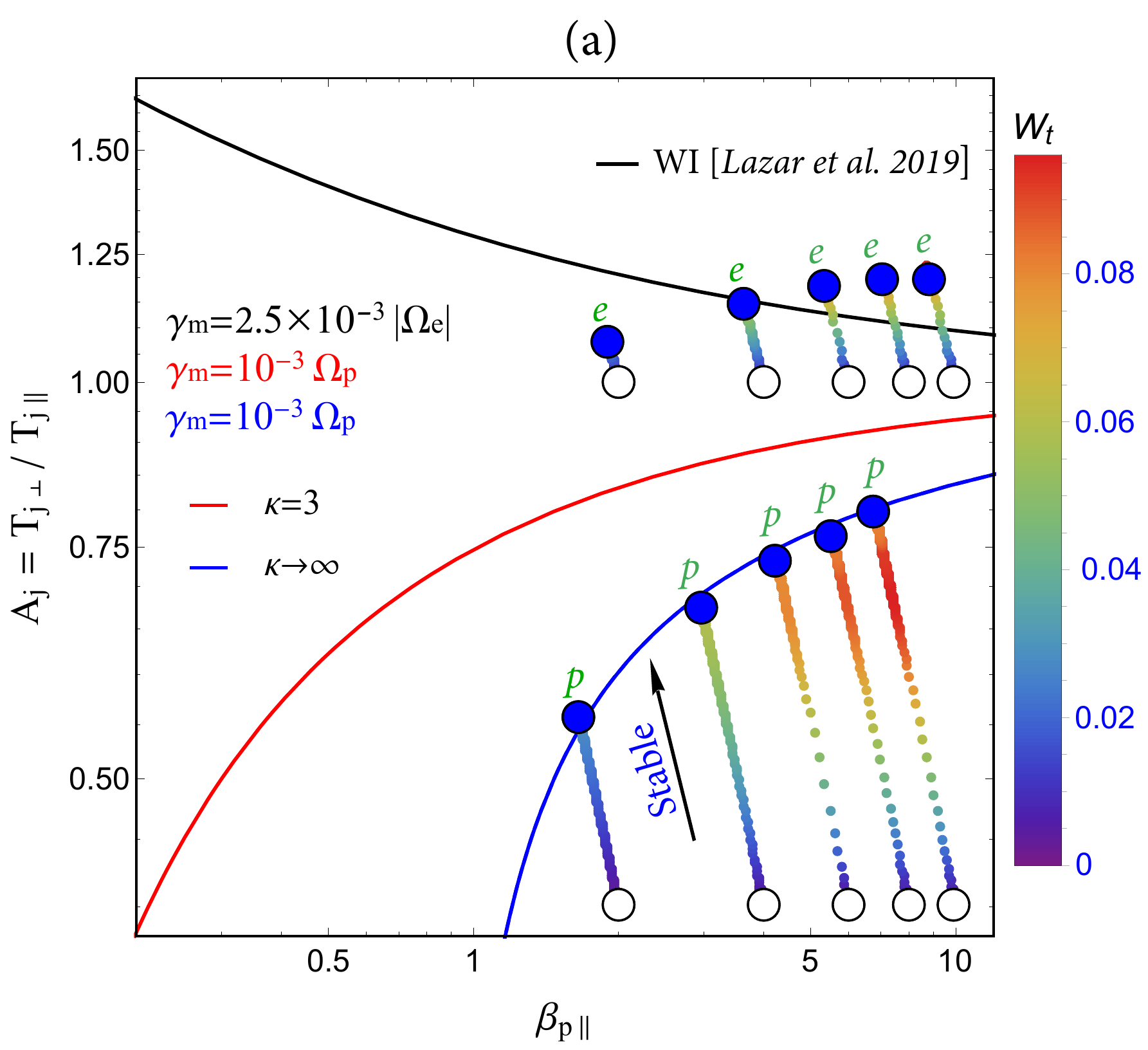}~~
    \includegraphics[scale=0.48]{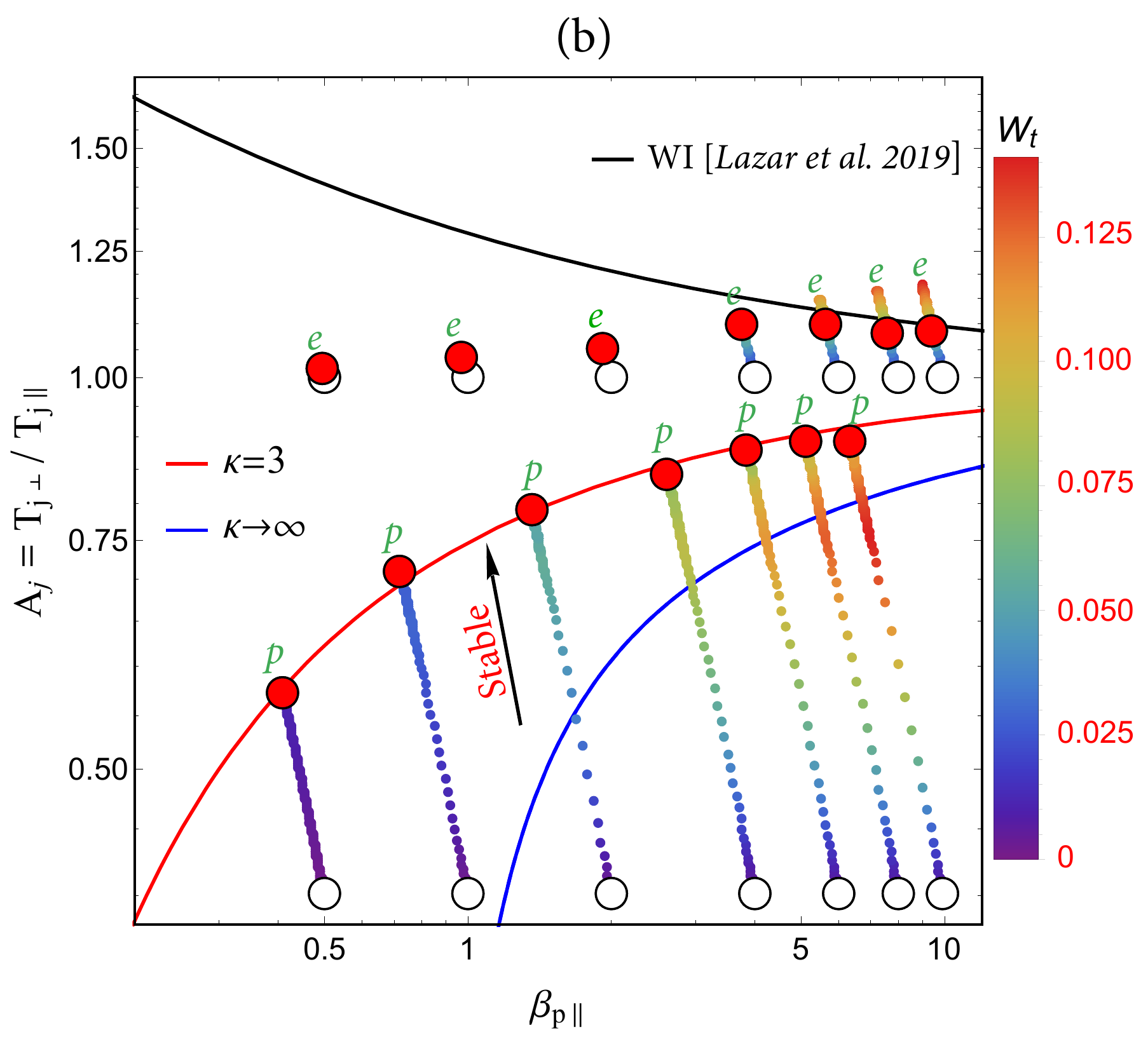}
    \caption{PFH instability thresholds derived from linear theory for $\kappa\rightarrow \infty$ (blue) $\kappa=3$ (red), and the QL dynamical decreasing paths for the proton and electrons in $(A_j, \beta_{p \parallel})-$space for $\kappa\rightarrow \infty$ (panel a) and $\kappa=3$ (panel b). The magnetic wave energy density level is colour coded. Initial states are indicated by white circles, while the final positions are marked with blue circles $\kappa\rightarrow \infty$ for and red circles for $\kappa=3$. Stable states of the PFH modes are indicated by black arrows.}
    \label{fig:6}
\end{figure*}
%

A direct consequence of the enhanced PFH fluctuations is the relaxation of the initial proton temperature anisotropy $A_p(0)<1$ through the wave-particle interaction, as already shown in Figure~\ref{fig:1}. The impact of the suprathermals on the temporal evolution of the proton and electron anisotropies can be illustrated in a more precise manner by calculating the instantaneous percent difference of the proton ($j=p$) and electron ($j=e$) temperature anisotropies $\Delta A_j (\tau)=~(A_j^{\kappa}(\tau)-~A_j^{M}(\tau))/A_j^{M}(\tau) \times ~100$, where $A_j^{\kappa}(\tau)$ and $A_j^{M}(\tau)$ are the instantaneous temperature anisotropies for the bi-Kappa (superscript $\kappa$) and bi-Maxwellian (superscript $M$) distributed protons, respectively.

\textcolor{black}{Figure~\ref{fig:5} displays the instantaneous percent difference $\Delta A_j (\tau)$ as a function of  $\tau$, for the proton (red) and electron (blue) temperature anisotropies. For $\beta_{p \parallel}=4$ (panel a) the percent difference of the proton temperature anisotropy shows exponential growth, and peaks at saturation ($\tau=43$) with value $\Delta A_{p, max}=63.8\%$ (and then the variation of anisotropy is less significant). In other words, the relaxation of the initial proton temperature anisotropy in the presence of suprathermal protons (i.e., for $\kappa=3$) is much stronger than that obtained for their bi-Maxwellian limit. For a higher proton beta parameter, i.e., for $\beta_{p \parallel}=6$ (panel b), peak value obtained at saturation is lower, i.e., $\Delta A_{p, max}(\tau)=34\%$. Electrons show the same behavior  in the presence of the suprathermal protons, with a percent difference of the electron temperature anisotropy peaking and then decreasing before $\tau_{max}$. However, for the electrons these peaks $\Delta A_{e, max}(\tau)$ are modest, and the saturated values are negative confirming the results in the right bottom panel of Figure~\ref{fig:1}, which show the initially isotropic electrons gaining less perpendicular anisotropy at the final stage $\tau_{max}$, i.e., $A^{M}_e(\tau_{max})>A^{\kappa}_e(\tau_{max})>1$. 
Panel (c) in Figure~\ref{fig:5} displays the variation of these two values of the percent difference of the proton temperature anisotropy with $\beta_{p \parallel}$. Bith the peaking value (red solid line) and the one obtained after saturation (red dashed line) decrease with  increasing $\beta_{p \parallel}$. Furthermore, the difference between $\Delta A_p(\tau)$ and $\Delta A_{p, max}(\tau_{max})$ decreases as $\beta_{p \parallel}$ increases, and for $\beta_{p \parallel}=10$, $\Delta A_p(\tau_{max})\simeq \Delta A_{p, max}(\tau)\simeq 15\%$.}

Figure~\ref{fig:6} displays the temperature anisotropy thresholds derived as a function of $\beta_{p \parallel}$  close to the marginal stability of PFH modes (low maximum growth rates $\gamma_{max}= 10^{-3} \Omega_p$), for $\kappa =3$ (red curves) and $\kappa \rightarrow \infty$ (blue curves). These thresholds are obtained from the linear dispersion relation \eqref{eq4} and are well fitted to \citep{Shaaban2017}
\begin{align}\label{eq12}
A_p=1-\frac{s}{\left(\beta_{p \parallel}-\beta_0\right)^{\alpha}}
\end{align}
with fitting parameters $(s, \alpha, \beta_0)= (0.45,0.47, 0.65)$ for $\kappa=3$ (red curve) and $(s, \alpha, \beta_0)=(0.26, 0.61,-0.042)$ for $\kappa \rightarrow \infty$ (blue).

In the $(A_j, \beta_{j \parallel})-$space the unstable PFH modes are located below the anisotropy thresholds, while the stable states are located above the thresholds, as indicted by the black arrows in Figure~\ref{fig:6}. These thresholds decrease with increasing $\beta_{p \parallel}$, extending the unstable regime of PFH modes to lower deviations from isotropy $A_p\lesssim 1$. This behavior is consistent with the fact that kinetic plasma modes need lower anisotropies to destabilize in hotter plasmas ($\beta_{p \parallel} \propto T_{p \parallel}$). \textcolor{black}{In Figure~\ref{fig:6} the effects of suprathermal protons are highlighted by a direct comparison of the anisotropy thresholds for $\kappa=3$ (red) and for bi-Maxwellian protons (blue). 
The anisotropy threshold becomes markedly lower in the presence of the suparthermal protons, extending the unstable regime of PFH modes to lower anisotropies and lower $\beta_{p \parallel}<1$. 
In Figure~\ref{fig:app} from Appendix A we display, in addition, the anisotropy threshold derived for the parallel plasma beta parameter of the bi-Kappa distributed protons (black line). This threshold  shows a similar significant displacement towards lower anisotropies and lower values of the plasma beta, due to the presence of suprathermal protons.}

Furthermore, with dots in Figure~\ref{fig:6} we display the results from QL approaches. We consider seven cases of distinct initial parameters by using different values for the parallel plasma beta parameter $\beta_{p \parallel}=0.5, 1, 2, 4, 6, 8$, and 10, and comparing again the results obtained for $\kappa=3$ and $\kappa \rightarrow \infty$. Other initial plasma parameters are $A_p(0)=0.3$, $A_e(0)=1.0$, and $\beta_{p \parallel}(0)=\beta_{e \parallel}(0)$. For all cases, the QL evolution of the  temperature anisotropy $A_j$ for protons (subscript "$j=p$") and electrons (subscript "$j=e$") are displayed as dynamical paths. The QL dynamical paths start at the initial conditions, as indicated by the white circles, and end at the final position after the saturation, as indicated by the red circles for $\kappa=3$ and the blue circles for $\kappa \to \infty$. \textcolor{black}{The level of the magnetic wave-energy density of the PFH fluctuations is indicated by the color bars, showing a clear enhancement in the presence of suprathermals (panel b).} The enhanced PFH fluctuations scatter protons towards the quasi-stable state close to marginal stability. Thus, for all cases the initial proton anisotropy reduces towards the quasi-stable states as time evolves, with a decline towards lower betas. \textcolor{black}{For all these QL runs the final states settle down exactly on the temperature anisotropy thresholds derived from linear theory. The presence of suprathermal protons, i.e., for $\kappa=3$ (panel b), determines a higher level of PFH fluctuations, and, in turn, a higher and more efficient relaxation of the proton temperature anisotropy. For bi-Maxwellian (core) protons (panel a), the PFH modes can be destabilized under the condition of $\beta_{p \parallel}\gtrsim 1$ \citep{Gary1993}, but in the presence of suprathermals this condition is markedly relaxed, requiring only $\beta_{p \parallel}>0.2$ (see red thresholds in Figure~\ref{fig:6}). Note, however, that this condition depends on the value chosen for $\kappa$, and the new minimum limit for $\beta_{p \parallel}$ increases with increasing $\kappa$.}

After gaining temperature anisotropy ($A_e>1$), the electrons may excite the RH polarized whistler instability (WI) with a maximum growth rate in the parallel direction to the background magnetic field \citep{Lazar2019}. For visual guidance Figure~\ref{fig:6} displays also the WI threshold (black curve) predicted from linear theory, with fitting parameters $(s,\alpha, \beta_0)=(-0.29, 0.49, 0.0)$ in Eq.~\eqref{eq12}. The dynamical paths of the initially isotropic electrons $A_e(0)$ show that electrons gain temperature anisotropy in the perpendicular direction and move toward the WI thresholds. In general, the electron temperature anisotropies $A_e$ induced after saturation are positioned below or near the WI threshold, except for $\beta_{p \parallel}>4$ in the case of the bi-Maxwellian distributed protons (panel a). This suggests that, despite the enhanced level of fluctuations in the presence of suprathermal protons, their resonant transfer of energy to electrons via the enhanced PFH fluctuations is reduced. Note, also, that an increase of the initial plasma beta parameter results in longer dynamical paths for the electrons and, implicitly, a higher gain of their induced  anisotropy.

\section{Conclusions}\label{Sec.4}
%
We have investigated the periodic PFH instability, in conditions typically encountered in space plasmas, where suprathermal particles are ubiquitous. To properly outline the effects of suparthermal protons we have performed a comparative analysis between the results obtained for bi-Kappa distributed protons, and those obtained in the absence of suprathermals, for the bi-Maxwellian (quasi-thermal) core \citep{Lazar2015Destabilizing, LazarAA2016}. Thus, Figures~\ref{fig:1}-\ref{fig:3} describe the effects of suprathermal protons on the linear properties, but also temporal evolution of PFH fluctuation and their back reactions on the plasma species, including macroscopic plasma parameters, i.e., plasma beta parameters $\beta_{j \perp, \parallel}$ and temperature anisotropies $A_j$, instantaneous growth rates of PFH instability, and the relaxation of the initial proton VDs. \textcolor{black}{All these results show a systematic stimulation of the instability in the presence of suprathermals, due to their additional free (kinetic) energy. Growth rates are enhanced, and so are the resulting PFH fluctuations, reaching higher levels of magnetic-wave energy density at the saturation. As a consequence of that, the relaxation of the proton temperature anisotropy becomes faster and more efficient, see Figures~\ref{fig:1} and \ref{fig:3}.}

\textcolor{black}{Figures~\ref{fig:4}-\ref{fig:6} provide a more comprehensive picture, showing the robustness of these stimulative effects of suprathermal protons on the PFH instability for an extended range of plasma beta conditions, in the interval $2\leqslant \beta_{p \parallel}\leqslant 10$. In  Figures~\ref{fig:4}-\ref{fig:5} the ignition time of PFH instability is markedly shortened in the presence of surprathermals, the maximum level of the enhanced fluctuations is enhanced, and the relaxation of anisotropic protons becomes more pronounced. Suprathermal protons contribute with an additional kinetic (free) energy (e.g., $T_\parallel^\kappa > T_\parallel$) and systematically stimulates the PFHI. In Figure~\ref{fig:6}, thresholds predicted by linear theory in a $(A_p, \beta_{p \parallel})$ diagram are exactly recovered from  the quasi-linear dynamical paths of the temperature anisotropy relaxation. The anisotropy thresholds are significantly reduced in the presence of suprathermal protons (i.e., $\kappa=3$), and the unstable regime is considerably expanded to lower beta regimes, i.e., $\beta_{p \parallel}<1$, where plasma dynamics is more constrained by the magnetic field. In this new regime the PFH modes are only unstable due to the free kinetic energy provided by suprathermal protons. Similar effects are induced by the suprathermal electrons on the conditions of electron firehose instabilities \citep{Lazar-etal-2017, Shaaban-etal-2019a}.}


To conclude, suprathermal protons have a significant and systematic stimulative effect on the PFH instability, in both linear and quasi-linear phases, enhancing not only the growth rates, but also the enhanced PFH fluctuations, which finally determine a faster instability development and a more efficient relaxation of the anisotropic protons. Comparing to idealized bi-Maxwellian plasmas, which completely ignore the effects of suprathermal protons, our results unveil a new unstable regime for the PFH instability, highly conditioned by the suprathermal protons and plasma beta parameter.

%
\section*{Acknowledgements}
%
The authors acknowledge support from the Katholieke Universiteit Leuven, Ruhr-University Bochum and Christian-Albrechts-Universit\"at Kiel. These results were obtained in the framework of the projects SCHL 201/35-1 (DFG-German Research Foundation), GOA/2015-014 (KU Leuven), G0A2316N (FWO-Vlaanderen). S.M.Shaaban acknowledges the Alexander-von-Humboldt Research Fellowship.
\textcolor{black}{\section*{Appendix A}
\begin{figure}
    \centering
    \includegraphics[width=0.35\textwidth]{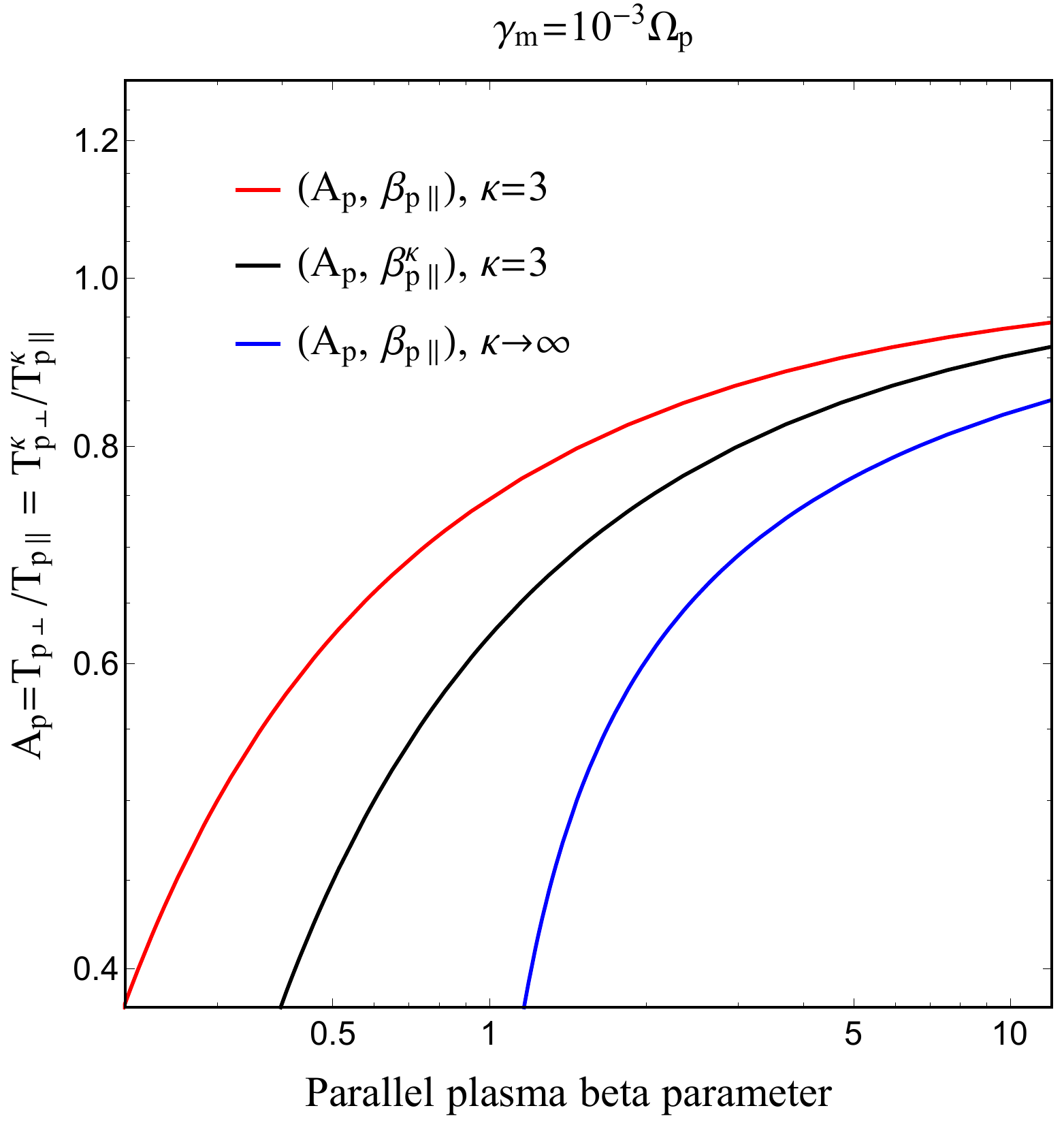}
    \caption{\textcolor{black}{A comparison of the PFH instability thresholds derived from linear theory in the $(A_p$, $\beta_{p \parallel})-$space for $\kappa\rightarrow \infty$ (blue), $\kappa=3$ (red), and in the $(A_p$, $\beta_{p \parallel}^\kappa)-$space for $\kappa=3$ (black).}}
    \label{fig:app}
\end{figure}
Figure~\ref{fig:app} presents a comparison of the PFH instability thresholds derived as a function of the parallel plasma beta parameter for $\kappa=3$ (red and black curves) and $\kappa \to \infty$ (blue). Red and blue curves are the same thresholds as in Figure~\ref{fig:6}, while the black curve represents the instability threshold in the $(A_p$, $\beta_{p \parallel}^\kappa)-$space, where $\beta_{p \parallel}^\kappa=2 \kappa/(2\kappa-3)\beta_{p \parallel}>\beta_{p \parallel}$ is the parallel plasma beta parameter for the (bi-)Kappa distributed protons. This threshold shows a similar significant displacement towards lower values of plasma beta, supporting the main conclusion of our present study.}

\bibliography{papers}

\begin{thebibliography}{}
\expandafter\ifx\csname natexlab\endcsname\relax\def\natexlab#1{#1}\fi
\providecommand{\url}[1]{\href{#1}{#1}}

\bibitem[{Astfalk \& Jenko(2016)}]{Astfalk2016}
Astfalk, P., \& Jenko, F. 2016, Journal of Geophysical Research A: Space
  Physics, 121, 2842.
\newblock \url{https://doi.org/10.1002/2015JA022267}

\bibitem[{Bale {et~al.}(2009)Bale, Kasper, Howes, Quataert, Salem, \&
  Sundkvist}]{Bale2009}
Bale, S., Kasper, J., Howes, G., {et~al.} 2009, Physical Review Letters, 103,
  211101.
\newblock \url{https://doi.org/10.1103/PhysRevLett.103.211101}

\bibitem[{Chew {et~al.}(1956)Chew, Goldberger, \& Low}]{Chew1956}
Chew, G.~F., Goldberger, M.~L., \& Low, F.~E. 1956, Proceedings of the Royal
  Society of London A: Mathematical, Physical and Engineering Sciences, 236,
  112.
\newblock \url{https://doi.org/10.1098/rspa.1956.0116}

\bibitem[{Christon {et~al.}(2017)Christon, Hamilton, Plane, Mitchell,
  Grebowsky, Spjeldvik, \& Nylund}]{Christon2017}
Christon, S.~P., Hamilton, D.~C., Plane, J. M.~C., {et~al.} 2017, Journal of
  Geophysical Research: Space Physics, 122, 11,175.
\newblock \url{https://doi.org/10.1002/2017JA024414}

\bibitem[{Christon {et~al.}(1989)Christon, Williams, Mitchell, Frank, \&
  Huang}]{Christon1989}
Christon, S.~P., Williams, D.~J., Mitchell, D.~G., Frank, L.~A., \& Huang,
  C.~Y. 1989, Journal of Geophysical Research: Space Physics, 94, 13409.
\newblock \url{https://doi.org/10.1029/JA094iA10p13409}

\bibitem[{Collier {et~al.}(1996)Collier, Hamilton, Gloeckler, Bochsler, \&
  Sheldon}]{Collier1996}
Collier, M.~R., Hamilton, D., Gloeckler, G., Bochsler, P., \& Sheldon, R. 1996,
  Geophysical Research Letters, 23, 1191.
\newblock \url{https://doi.org/10.1029/96GL00621}

\bibitem[{Davidson \& Völk(1968)}]{Davidson1968}
Davidson, R.~C., \& Völk, H.~J. 1968, The Physics of Fluids, 11, 2259.
\newblock \url{https://doi.org/10.1063/1.1691810}

\bibitem[{{Ebert} {et~al.}(2012){Ebert}, {Dayeh}, {Desai}, \&
  {Mason}}]{Ebert2012}
{Ebert}, R.~W., {Dayeh}, M.~A., {Desai}, M.~I., \& {Mason}, G.~M. 2012, The
  Astrophysical Journal, 749, 73.
\newblock \url{https://doi.org/10.1088/0004-637X/749/1/73}

\bibitem[{Fried \& Conte(1961)}]{Fried1961}
Fried, B., \& Conte, S. 1961, The Plasma Dispersion Function (New York:
  Academic Press).
\newblock \url{https://doi.org/10.1016/B978-1-4832-2929-4.50005-8}

\bibitem[{Gary(1993)}]{Gary1993}
Gary, S.~P. 1993, Theory of Space Plasma Microinstabilities (Cambridge
  university press).
\newblock \url{https://doi.org/10.1017/CBO9780511551512}

\bibitem[{Hellinger {et~al.}(2006)Hellinger, Tr{\'a}vn{\'\i}{\v{c}}ek, Kasper,
  \& Lazarus}]{Hellinger2006}
Hellinger, P., Tr{\'a}vn{\'\i}{\v{c}}ek, P., Kasper, J.~C., \& Lazarus, A.~J.
  2006, Geophysical Research Letters, 33.
\newblock \url{https://doi.org/10.1029/2006GL025925}

\bibitem[{Huang {et~al.}(2020)Huang, Kasper, Vech, Klein, Stevens,
  Martinovi{\'{c}}, Alterman, {\v{D}}urovcov{\'{a}}, Paulson, Maruca, Qudsi,
  Case, Korreck, Jian, Velli, Lavraud, Hegedus, Bert, Holmes, Bale, Larson,
  Livi, Whittlesey, Pulupa, MacDowall, Malaspina, Bonnell, Harvey, Goetz, \&
  de~Wit}]{Huang2020}
Huang, J., Kasper, J.~C., Vech, D., {et~al.} 2020, The Astrophysical Journal
  Supplement Series, 246, 70.
\newblock \url{https://doi.org/10.3847/1538-4365/ab74e0}

\bibitem[{Hunana \& Zank(2017)}]{Hunana2017}
Hunana, P., \& Zank, G.~P. 2017, The Astrophysical Journal, 839, 13.
\newblock \url{https://doi.org/10.3847/1538-4357/aa64e3}

\bibitem[{Jeong {et~al.}(2020)Jeong, Verscharen, Wicks, \&
  Fazakerley}]{Jeong2020}
Jeong, S.-Y., Verscharen, D., Wicks, R.~T., \& Fazakerley, A.~N. 2020, The
  Astrophysical Journal, 902, 128.
\newblock \url{https://doi.org/10.3847/1538-4357/abb099}

\bibitem[{Kasper {et~al.}(2002)Kasper, Lazarus, \& Gary}]{Kasper2002}
Kasper, J.~C., Lazarus, A.~J., \& Gary, S.~P. 2002, Geophysical Research
  Letters, 29.
\newblock \url{https://doi.org/10.1029/2002GL015128}

\bibitem[{Kasper {et~al.}(2003)Kasper, Lazarus, Gary, \& Szabo}]{Kasper2003}
Kasper, J.~C., Lazarus, A.~J., Gary, S.~P., \& Szabo, A. 2003, AIP Conference
  Proceedings, 679, 538.
\newblock \url{https://doi.org/10.1063/1.1618653}

\bibitem[{Kim {et~al.}(2018)Kim, Lazar, Schlickeiser, L{\'{o}}pez, \&
  Yoon}]{Kim2018}
Kim, S., Lazar, M., Schlickeiser, R., L{\'{o}}pez, R.~A., \& Yoon, P.~H. 2018,
  Plasma Physics and Controlled Fusion, 60, 075010.
\newblock \url{https://doi.org/10.1088/1361-6587/aac1e4}

\bibitem[{Lario {et~al.}(2019)Lario, Berger, Decker, Wimmer-Schweingruber,
  Wilson, Giacalone, \& Roelof}]{Lario2019}
Lario, D., Berger, L., Decker, R.~B., {et~al.} 2019, The Astronomical Journal,
  158, 12.
\newblock \url{https://doi.org/10.3847/1538-3881/ab1e49}

\bibitem[{{Lazar} {et~al.}(2016){Lazar}, {Fichtner}, \& {Yoon}}]{LazarAA2016}
{Lazar}, M., {Fichtner}, H., \& {Yoon}, P.~H. 2016, \aap, 589, A39.
\newblock \url{https://doi.org/10.1051/0004-6361/201527593}

\bibitem[{Lazar {et~al.}(2018{\natexlab{a}})Lazar, Kim, L{\'{o}}pez, Yoon,
  Schlickeiser, \& Poedts}]{Lazar2018Spont}
Lazar, M., Kim, S., L{\'{o}}pez, R.~A., {et~al.} 2018{\natexlab{a}}, The
  Astrophysical Journal Letters, 868, L25.
\newblock \url{https://doi.org/10.3847/2041-8213/aaefec}

\bibitem[{Lazar {et~al.}(2019)Lazar, L{\'{o}}pez, Shaaban, Poedts, \&
  Fichtner}]{Lazar2019}
Lazar, M., L{\'{o}}pez, R.~A., Shaaban, S.~M., Poedts, S., \& Fichtner, H.
  2019, Astrophysics and Space Science, 364, 171.
\newblock \url{http://doi.org/10.1007/s10509-019-3661-6}

\bibitem[{{Lazar} {et~al.}(2015){Lazar}, {Poedts}, \&
  {Fichtner}}]{Lazar2015Destabilizing}
{Lazar}, M., {Poedts}, S., \& {Fichtner}, H. 2015, \aap, 582, A124.
\newblock \url{https://doi.org/10.1051/0004-6361/201526509}

\bibitem[{Lazar {et~al.}(2011)Lazar, Poedts, \& Schlickeiser}]{Lazar2011}
Lazar, M., Poedts, S., \& Schlickeiser, R. 2011, \aap, 534, A116.
\newblock \url{https://doi.org/10.1051/0004-6361/201116982}

\bibitem[{{Lazar} {et~al.}(2008){Lazar}, {Schlickeiser}, \&
  {Shukla}}]{Lazar2008}
{Lazar}, M., {Schlickeiser}, R., \& {Shukla}, P.~K. 2008, Physics of Plasmas
  (1994-present), 15, 042103.
\newblock \url{https://doi.org/10.1063/1.2896232}

\bibitem[{{Lazar} {et~al.}(2017){Lazar}, {Shaaban}, {Poedts}, \&
  {{\v{S}}tver{\'a}k}}]{Lazar-etal-2017}
{Lazar}, M., {Shaaban}, S.~M., {Poedts}, S., \& {{\v{S}}tver{\'a}k}, {\v{S}}.
  2017, \mnras, 464, 564.
\newblock \url{https://doi.org/10.1093/mnras/stw2336}

\bibitem[{Lazar {et~al.}(2018{\natexlab{b}})Lazar, Yoon, López, \&
  Moya}]{Lazar2018JGR}
Lazar, M., Yoon, P.~H., López, R.~A., \& Moya, P.~S. 2018{\natexlab{b}},
  Journal of Geophysical Research: Space Physics, 123, 6.
\newblock \url{https://doi.org/10.1002/2017JA024759}

\bibitem[{Maneva {et~al.}(2016)Maneva, Lazar, Vi{\~{n}}as, \&
  Poedts}]{Maneva2016}
Maneva, Y., Lazar, M., Vi{\~{n}}as, A., \& Poedts, S. 2016, The Astrophysical
  Journal, 832, 64.
\newblock \url{https://doi.org/10.3847/0004-637x/832/1/64}

\bibitem[{Marsch(2006)}]{Marsch2006}
Marsch, E. 2006, Living Reviews in Solar Physics, 3, 1.
\newblock \url{https://doi.org/10.12942/lrsp-2006-1}

\bibitem[{Matteini {et~al.}(2007)Matteini, Landi, Hellinger, Pantellini,
  Maksimovic, Velli, Goldstein, \& Marsch}]{Matteini2007}
Matteini, L., Landi, S., Hellinger, P., {et~al.} 2007, Geophysical Research
  Letters, 34, L20105.
\newblock \url{https://doi.org/10.1029/2007GL030920}

\bibitem[{Micera {et~al.}(2020)Micera, Boella, Zhukov, Shaaban, L{\'{o}}pez,
  Lazar, \& Lapenta}]{Micera2020}
Micera, A., Boella, E., Zhukov, A.~N., {et~al.} 2020, The Astrophysical
  Journal, 893, 130.
\newblock \url{http://dx.doi.org/10.3847/1538-4357/ab7faa}

\bibitem[{Michno {et~al.}(2014)Michno, Lazar, Yoon, \&
  Schlickeiser}]{Michno2014}
Michno, M., Lazar, M., Yoon, P., \& Schlickeiser, R. 2014, The Astrophysical
  Journal, 781, 49.
\newblock \url{https://doi.org/10.1088/0004-637X/781/1/49}

\bibitem[{Moya {et~al.}(2020)Moya, Lazar, \& Poedts}]{Moya2021}
Moya, P.~S., Lazar, M., \& Poedts, S. 2020, Plasma Physics and Controlled
  Fusion, 63, 025011.
\newblock \url{https://doi.org/10.1088/1361-6587/abce1a}

\bibitem[{Pierrard \& Lazar(2010)}]{Pierrard2010}
Pierrard, V., \& Lazar, M. 2010, Solar Physics, 267, 153.
\newblock \url{https://doi.org/10.1007/s11207-010-9640-2}

\bibitem[{Seough {et~al.}(2014)Seough, Yoon, \& Hwang}]{Seough2014}
Seough, J., Yoon, P.~H., \& Hwang, J. 2014, Physics of Plasmas, 21, 062118.
\newblock \url{https://doi.org/10.1063/1.4885359}

\bibitem[{Seough {et~al.}(2015)Seough, Yoon, \& Hwang}]{Seough2015}
---. 2015, Physics of Plasmas, 22, 012303.
\newblock \url{https://doi.org/10.1063/1.4905230}

\bibitem[{Shaaban {et~al.}(2017)Shaaban, Lazar, Poedts, \&
  Elhanbaly}]{Shaaban2017}
Shaaban, S., Lazar, M., Poedts, S., \& Elhanbaly, A. 2017, Astrophysics and
  Space Science, 362, 13.
\newblock \url{https://doi.org/10.1007/s10509-016-2994-7}

\bibitem[{{Shaaban} {et~al.}(2019a){Shaaban}, {Lazar}, {L{\'o}pez}, {Fichtner},
  \& {Poedts}}]{Shaaban-etal-2019a}
{Shaaban}, S.~M., {Lazar}, M., {L{\'o}pez}, R.~A., {Fichtner}, H., \& {Poedts},
  S. 2019a, \mnras, 483, 5642.
\newblock \url{https://doi.org/10.1093/mnras/sty3377}

\bibitem[{Shaaban {et~al.}(2021{\natexlab{a}})Shaaban, Lazar, L{\'{o}}pez, \&
  Wimmer-Schweingruber}]{ShaabanMNRAS2021}
Shaaban, S.~M., Lazar, M., L{\'{o}}pez, R.~A., \& Wimmer-Schweingruber, R.~F.
  2021{\natexlab{a}}, \mnras, 503, 3134.
\newblock \url{https://doi.org/10.1093/mnras/stab075}

\bibitem[{Shaaban {et~al.}(2016)Shaaban, Lazar, Poedts, \&
  Elhanbaly}]{Shaaban2016Supra}
Shaaban, S.~M., Lazar, M., Poedts, S., \& Elhanbaly, A. 2016, Astrophysics and
  Space Science, 361, 1.
\newblock \url{https://doi.org/10.1007/s10509-016-2782-4}

\bibitem[{Shaaban {et~al.}(2021{\natexlab{b}})Shaaban, Lazar, \&
  Schlickeiser}]{Shaaban2021}
Shaaban, S.~M., Lazar, M., \& Schlickeiser, R. 2021{\natexlab{b}}, Physics of
  Plasmas, 28, 022103.
\newblock \url{https://doi.org/10.1063/5.0035798}

\bibitem[{{Shaaban} {et~al.}(2019b){Shaaban}, {Lazar}, {Yoon}, \&
  {Poedts}}]{ShaabanApJ2019}
{Shaaban}, S.~M., {Lazar}, M., {Yoon}, P.~H., \& {Poedts}, S. 2019b, \apj, 871,
  237.
\newblock \url{https://doi.org/10.3847/1538-4357/aaf72d}

\bibitem[{{\v{S}}tver{\'a}k {et~al.}(2008){\v{S}}tver{\'a}k,
  Tr{\'a}vn{\'\i}{\v{c}}ek, Maksimovic, Marsch, Fazakerley, \&
  Scime}]{Stverak2008}
{\v{S}}tver{\'a}k, {\v{S}}., Tr{\'a}vn{\'\i}{\v{c}}ek, P., Maksimovic, M.,
  {et~al.} 2008, Journal of Geophysical Research: Space Physics, 113.
\newblock \url{https://doi.org/10.1029/2007JA012733}

\bibitem[{Summers \& Thorne(1991)}]{Summers1991}
Summers, D., \& Thorne, R.~M. 1991, Physics of Fluids B: Plasma Physics
  (1989-1993), 3, 1835.
\newblock \url{https://doi.org/10.1063/1.859653}

\bibitem[{{Tylka} \& {Lee}(2006)}]{Tylka2006}
{Tylka}, A.~J., \& {Lee}, M.~A. 2006, The Astrophysical Journal, 646, 1319.
\newblock \url{https://doi.org/10.1086/505106}

\bibitem[{{Yoon}(2017)}]{Yoon2017}
{Yoon}, P.~H. 2017, Reviews of Modern Plasma Physics, 1, 4.
\newblock \url{https://doi.org/10.1007/s41614-017-0006-1}

\bibitem[{Yoon {et~al.}(2017)Yoon, López, Seough, \& Sarfraz}]{Yoon2017PoP}
Yoon, P.~H., López, R.~A., Seough, J., \& Sarfraz, M. 2017, Physics of
  Plasmas, 24, 112104.
\newblock \url{https://doi.org/10.1063/1.4997666}

\bibitem[{Yu {et~al.}(2017)Yu, Berger, Wimmer-Schweingruber, Bochsler, Klecker,
  Hilchenbach, \& Kallenbach}]{Yu2017}
Yu, J., Berger, L., Wimmer-Schweingruber, R., {et~al.} 2017, \aap, 599, 1.
\newblock \url{http://doi.org/10.1051/0004-6361/201628641}

\end{thebibliography}

%
\listofchanges
\end{document}